\documentstyle[12pt]{article}
\begin{document}

\title{\bf\huge 
Bethe ansatz for the one-dimensional small-polaron model with open
boundary conditions
}

\author{
{\bf
Heng Fan$^{a,b}$, Xi-Wen Guan$^{a,c}$}\\
\normalsize $^a$ CCAST(World Laboratory)\\
\normalsize P.O.Box 8730,Beijing 100080,China\\
\normalsize $^b$ Institute of Modern Physics,Northwest University\\
\normalsize P.O.Box 105,Xian,710069,China\thanks{Mailing address}\\
\normalsize $^c$ Center for Theoretical Physics and Department of Physics,\\
\normalsize Jilin University, Changchun 130023, China
}
\maketitle

\begin{abstract}
The one-dimensional 
small-polaron model with open boundary conditions is
considered in the framework of the quantum inverse scattering method. 
The spin model which is equivalent to the small-polaron
model is the Heisenberg $XXZ$ spin chain 
in an external magnetic field. 
For spin model,                                           
the solutions of the reflection equation (RE) and the dual RE are
obtained. The eigenvalues and
eigenvectors problems are solved by using the algebraic Bethe ansatz 
method. The case of the fermion model is also studied, the commuting
of the transfer matrix can be proved, and the eigenvalues
and the Bethe ansatz equations for fermion model are obtained.
\end{abstract}
\vskip 1truecm
PACS: 7510Jm, 0520, 0530.

\noindent Keywords: One-dimensional small-polaron model,
Heisenberg $XXZ$ spin chain, Yang-Baxter eqution, reflection equation.

\newpage
\section{Introduction}
Recently, much more attention has been paid to study the
integrable models with open boundary conditions [1-10].
This was initiated by Sklyanin [1], who proposed a systematic approach
to handle integrable models with reflecting boundary conditions in
which the so-called RE plays a key role.

Sklyanin assume that the R-matrix of the vertex models are
$P$ and $T$ symmetric $R_{12}=PR_{12}P=R_{21},
R_{12}^{t_1t_2}=R_{12}$, where $t_i, i=1,2$ means transposition in
the $i$-th space, $P$ is the permutation operator
$Px\otimes y=y\otimes x$. Because only few models
satisfy these properties, Mezincescu $et~al$ [3] extended Sklyanin's
formalism for constructing integrable open chains to the case where
the R matrix is only $PT$-invariant
$PR_{12}P=R^{t_1t_2}_{12}$. It is later realized that 
the unitarity and cross-unitarity relations of the R-matrix
are enough to prove
integrable open boundary conditions. 
In the framework of the quantum inverse scattering method (QISM) [11],
a lot of models with open boundary conditions are solved by using the 
algebraic Bethe ansatz method [2-7]. 
And the boundary cross-unitarity is also proposed for
open boundary conditions problems [8].

In this paper, we will study the one-dimensional
small-polaron model which
describes the motion of an additional electron in a polar crystal [12].
Via the Jordan-Wigner transformation, 
the one-dimensional small-polaron 
model is equivalent to the
Heisenberg $XXZ$ spin chain with an external magnetic field
parallel to the $z$ direction [13]. We will consider 
the open boundary conditions for this
Heisenberg $XXZ$ spin chain. 
The R-matrix satisfy the unitarity and the cross-unitarity 
relations, so we can formulate the RE and 
the dual RE. With the help of these relations 
we can prove the
commuting of the transfer matrix with open boundary conditions.
The commuting transfer matrix can
give an infinite set of conserved quantities,
it is general in this sense that the model under consideration
is integrable. 
It is well known that 
the first non-trivial conserved quantities is the
Hamiltonian. 
Solving the RE and the dual RE, we find the diagonal solutions 
$K$ matrices of those REs. 
From the transfer matrix with open boundary conditions,
we obtain the Hamiltonin which includes
the boundary terms in the
Heisenberg $XXZ$ spin chain with an external 
gnetic field parallel to the $z$ direction.
Those boundary terms are determined by the boundary $K$ matrices.
Using the algebraic Bethe ansatz method, we find
the eigenvalues and eigenvectors of the transfer matrix. So the
eigenvalues of the Hamiltonian can also be obtained.

Using similar procedure, the one-dimensional small-polaron model
with open boundary conditions is studied. The eigenvalues of
the transfer matrix are obtained by using the algebraic Bethe ansatz
method.

The paper is organized as follows: In section 2, we will introduce the
model and formulate the RE and dual RE. We then prove the commuting
of the transfer matrix with open boundary conditions in section 3.
In section 4, we use the algebraic Bethe ansatz method to solve
the eigenvalues and eigenvectors problems. Section 5 will be devoted 
to the fermion case, and the eigenvalues and the corresponding
Bethe ansatz equations are obtained for the one-dimensional
small-polaron model with open boundary conditions. Finally, we 
will give a summary and discussions.

\section{The model and the RE and dual RE}
\indent As mentioned in the introduction, the one-dimensional
small-polaron model describes the motion of an additional
electron in a polar crystal [12]. The Hamiltonian takes the
form
\begin{eqnarray}
{\cal {H}}=W\sum _{j=1}^Nn_j
-J\sum _{j=1}^N(a_j^{\dagger }a_{j+1}+a_{j+1}^{\dagger }a_j)
+V\sum _{j=1}^Nn_jn_{j+1},
\end{eqnarray}
where $a_j^{\dagger }, a_j$ are fermion creation and 
annihilation operators at lattice site $j$, and satisfy the 
anticommutation relations
\begin{eqnarray}
\{ a_i, a_j\} =\{ a_i^{\dagger }, a_j^{\dagger }\} =0,
\{ a_i, a_j^{\dagger }\} =\delta _{ij},
\end{eqnarray}
$n_j=a_j^{\dagger }a_j$ is the occupation number operator.
$W, J$ and $V$ can be found in the paper of Fedyanin $et~al$ [12].

Applying the the Jordan-Wigner transformation for $a_j, a_j^{\dagger }$ and
$n_j$ 
\begin{eqnarray}
a_j&=&{\rm exp}
\left( i\pi \sum _{l=1}^{j-1}\sigma _l^+\sigma _l^-\right)
\sigma _j^-, \nonumber \\
a_j^{\dagger }&=&{\rm exp}\left( i\pi \sum _{l=1}^{j-1}\sigma _l^+
\sigma _l^-\right) \sigma _j^+, \nonumber \\
n_j&=&\frac {(1+\sigma _j^z)}{2},
\end{eqnarray}
where $\sigma _j^{\pm }={1\over 2}(\sigma _j^x\pm i\sigma _j^y)$ and
$\sigma _j^x, \sigma _j^y, \sigma _j^z$ are Pauli operators at lattice
site $j$. The Hamiltonian of the one-dimensional small-polaron model
changes to the Heisenberg $XXZ$ spin chain with an external magnetic
field parallel to $z$ direction.
\begin{eqnarray}
H=-\sum _{j=1}^N\left( J(\sigma _j^+\sigma _{j+1}^-
+\sigma _j^-\sigma _{j+1}^+)-{V\over 4}
\sigma _j^z\sigma _{j+1}^z\right)
+{{W+V}\over 2}\sum _{j=1}^N\sigma _j^z+cost..
\end{eqnarray}
Here the model is considered in the periodic boundary conditions.

From this Hamilotanian, 
the local monodromy matrix $L$ operator is found to be [13]
\begin{eqnarray}
L_j(\lambda )=\left(
\begin{array}{cc}
\frac {a'_++b'_+}{2}+\frac {a'_+-b'_+}{2}\sigma _j^z
& c\sigma _j^-\\
c\sigma _j^+ & \frac {a'_-+b'_-}{2}-\frac {a'_--b'_-}{2}\sigma _j^z
\end{array}\right),
\end{eqnarray}
here we have used the notations
\begin{eqnarray}
a_+':b_+':a_-':b_-':c
&=&\xi (\lambda )\sin (\lambda +2\eta )
:\xi ^{-1}(\lambda )\sin (\lambda )
\nonumber \\
&&: \xi ^{-1}(\lambda )\sin (\lambda +2\eta ):
\xi (\lambda )\sin (\lambda ):\sin (2\eta ),
\end{eqnarray}
with 
\begin{eqnarray}
\xi (\lambda )=\sec (\alpha )\cos (\lambda +\alpha )\sec (\lambda ). 
\end{eqnarray}
And $J, V$ and $W$ are given by 
\begin{eqnarray}
J:-{V\over 2}:(W+V)=1:\cos (2\eta ):2\sin (2\eta )\tan (\alpha ).
\end{eqnarray}

It is well known that for periodic boundary condition problems,
the Yang-Baxter relation [14,15]
plays an essential role which takes the form
\begin{equation}
R_{12}(\lambda ,\mu)L_1(\lambda )
L_2(\mu )=L_2(\mu )L_1(\lambda )R_{12}(\lambda ,\mu),
\end{equation}
\noindent where the indices of $L$ represent the auxiliary spaces, and
the R-matrix can be found to be:
\begin{eqnarray}
R_{12}(\lambda ,\mu )
\equiv \left(
\begin{array}{cccc}
a_+(\lambda ,\mu ) & 0& 0& 0\\
0 &b_-(\lambda ,\mu )& c(\lambda ,\mu )& 0\\
0 &c(\lambda ,\mu )& b_+(\lambda ,\mu )& 0\\
0& 0& 0& a_-(\lambda ,\mu )
\end{array} \right) , 
\end{eqnarray}
the non-zero elements of the R-matrix are defined as
\begin{eqnarray}
a_+(\lambda ,\mu )=\xi (\lambda )\xi ^{-1}(\mu )\sin (\lambda -\mu
-2\eta ), ~~
&&a_-(\lambda ,\mu )=\xi ^{-1}(\lambda )\xi (\mu )\sin (\lambda -\mu
-2\eta ), \nonumber\\
b_-(\lambda ,\mu )=\xi ^{-1}(\lambda )\xi ^{-1}(\mu )\sin (\lambda -\mu ),
&&
b_+(\lambda ,\mu )=\xi (\lambda )\xi (\mu )\sin (\lambda -\mu ),
\nonumber \\
c(\lambda ,\mu )=\sin (2\eta ),
\end{eqnarray}
here $\eta $ is the anisotropic parameter. 
In the framework of the QISM,
The row-to-row monodromy matrix is defined as:
\begin{eqnarray}
T(\lambda )=L_N(\lambda )\cdots L_1(\lambda ) 
\end{eqnarray}      
Here we know that the indices of the operator $L$
mean the quantum spaces. 
It is known that using repeatedly the Yang-Baxter relation (9), we
have
\begin{eqnarray}
R_{12}(\lambda ,\mu)T_1(\lambda )
T_2(\mu )=T_2(\mu )T_1(\lambda )R_{12}(\lambda ,\mu).
\end{eqnarray}
For periodic boundary conditions, we define the transfer matrix as
trace of monodromy matrix $T$ in the auxiliary space. Here,
to avoid confusion,
we represent the auxiliary space by $\bar {0}$.
\begin{eqnarray}
t_{peri.}(\lambda )&=&tr_{\bar {0}}T_{\bar {0}}(\lambda )\nonumber \\
&=&tr_{\bar {0}}L_{N{\bar {0}}}(\lambda )
\cdots L_{1{\bar {0}}}(\lambda ).
\end{eqnarray}
The Hamiltonian (4) can
be obtained from this transfer matrix 
\begin{eqnarray}
H=-\sin (2\eta )\frac {d}{d\lambda }\ln t_{peri.}(\lambda )|_{\lambda =0}.
\end{eqnarray}

What we study
in this paper is the reflecting open boundary conditions.
For open boundary condition case, besides the Yang-Baxter relation,
we also need the RE and the dual RE. Generally, we write the RE as
the following form
\begin{eqnarray}
R_{12}(\lambda ,\mu )K_1(\lambda )R_{21}(\mu ,-\lambda )K_2(\mu ) 
=K_2(\mu )R_{12}(\lambda ,-\mu )K_1(\lambda )R_{21}(-\mu ,-\lambda ).
\end{eqnarray}
The form of the dual RE depends on the unitarity and the cross-unitarity
relations of the R-matrix.  For the model considering in this paper, the
unitarity and cross-unitarity relations of the R-matrix are written as
\begin{eqnarray}
R_{12}(\lambda ,\mu )R_{21}(\mu ,\lambda )
=\sin (2\eta +\lambda -\mu )\sin (2\eta -\lambda +\mu )
\equiv \rho (\lambda -\mu ),
\\
X_{12}^{t_1}(\lambda ,\mu +4\eta )R_{21}^{t_1}(\mu ,\lambda )
=-\sin (\lambda -\mu )\sin (\lambda -\mu -4\eta )
\equiv \tilde {\rho }(\lambda -\mu ).
\end{eqnarray}
Here, for convenience, we have introduced the notations
\begin{eqnarray}
X_{12}(\lambda ,\mu )\equiv
\frac {\xi (\mu )}{\xi (\mu -4\eta )}M_1(\mu )
R_{12}(\lambda ,\mu )M_1(\mu ),
\nonumber \\
M(\lambda )=\left( \begin{array}{cc}
1 & 0\\ 
0 &\xi (\lambda -4\eta )\xi ^{-1}(\lambda )\end{array}\right) .
\end{eqnarray}
It is also convenient to write here another form of the cross-unitarity
relation:
\begin{eqnarray}
&&R_{12}^{t_1}(\lambda ,\mu )Y_{21}^{t_1}(\mu -4\eta ,\lambda )
=\tilde {\rho }(\lambda -\mu +4\eta ),
\nonumber
\\
&&Y_{12}(\lambda ,\mu )=\frac {\xi (\lambda +4\eta )}{\xi (\lambda )}
M_2(\lambda )R_{12}(\lambda , \mu )M_2(\lambda ).
\end{eqnarray}
Thus, we take the dual RE as the following form:
\begin{eqnarray}
R_{12}(-\lambda ,-\mu )K^+_1(\lambda )
Y_{21}(-\mu -4\eta ,\lambda )K_2^+(\mu )
\nonumber 
\\
=K_2^+(\mu )X_{12}(-\lambda ,\mu +4\eta )K_1(\lambda )
R_{21}(\mu ,\lambda ).
\end{eqnarray}
Following the method of Sklyanin, we define the double-row-monodromy
matrix as:
\begin{eqnarray}
{\cal {T}}(\lambda )=T(\lambda )K(\lambda )T^{-1}(-\lambda ).
\end{eqnarray}
Using the Yang-Baxter relation (13) and the RE, it can be proved that
the double-row-monodromy matrix also satisfy the RE:
\begin{eqnarray}
R_{12}(\lambda ,\mu ){\cal {T}}_1(\lambda )
R_{21}(\mu ,-\lambda ){\cal {T}}_2(\mu ) 
={\cal {T}}_2(\mu )R_{12}(\lambda ,-\mu )
{\cal {T}}_1(\lambda )R_{21}(-\mu ,-\lambda ).
\end{eqnarray}
We define the transfer matrix for open boundary conditions as:
\begin{eqnarray}
t(\lambda )=trK^+(\lambda ){\cal {T}}(\lambda ).
\end{eqnarray}

\section{Integrable open boundary conditions}

In this section, we will prove that the transfer matrix defined
above for open boundary conditions constitute a commuting family.
And we know that it is generally in this sense we mean the model
is integrable. Now, let us prove the transfer matrix commute for
different spectral parameters.
\begin{eqnarray}
t(\lambda )t(\mu )&=&tr_1K_1^+(\lambda ){\cal {T}}_1(\lambda )
tr_2K_2^+(\mu ){\cal {T}}_2(\mu )
\nonumber \\
&=&tr_{12}K_1^+(\lambda )^{t_1}K_2^+(\mu ){\cal {T}}_1^{t_1}(\lambda )
{\cal {T}}_2(\mu ) ,\nonumber 
\end{eqnarray}
here we take transpositon $t_1$ in space 1. Now we insert the
cross-unitrarity
relation (18), we have
\begin{eqnarray}
\cdots &=&\frac {1}{\tilde {\rho }(-\lambda -\mu )}
tr_{12}K_1^+(\lambda )^{t_1}K_2^+(\mu )
X_{12}^{t_1}(-\lambda ,\mu +4\eta )R_{21}^{t_1}(\mu ,-\lambda )
{\cal {T}}_1^{t_1}(\lambda )
{\cal {T}}_2(\mu ) ,\nonumber \\
&=&\frac {1}{\tilde {\rho }(-\lambda -\mu )}
tr_{12}\{ K_1^+(\lambda )^{t_1}K_2^+(\mu )
X_{12}^{t_1}(-\lambda ,\mu +4\eta )\} ^{t_1}
\{ {\cal {T}}_1(\lambda )R_{21}(\mu ,-\lambda )
{\cal {T}}_2(\mu ) \},\nonumber \\
&=&\frac {1}{\tilde {\rho }(-\lambda -\mu )}
tr_{12}\{ K_2^+(\mu )
X_{12}(-\lambda ,\mu +4\eta )K_1^+(\lambda )
\} \{ {\cal {T}}_1(\lambda )R_{21}(\mu ,-\lambda )
{\cal {T}}_2(\mu ) \} .
\nonumber 
\end{eqnarray}
In the above calculation, we take transposition in the first space.
Insert the untarity relation of the R-matrix (17), and use the RE (23) 
and the dual RE (21), we have
\begin{eqnarray}
\cdots &=&\frac {1}{\tilde {\rho }(-\lambda -\mu )\rho (\lambda -\mu )}
tr_{12}\{ K_2^+(\mu )
X_{12}(-\lambda ,\mu +4\eta )K_1^+(\lambda )
R_{21}(\mu ,\lambda )\} \nonumber \\
&&~~~\{ R_{12}(\lambda ,\mu ){\cal {T}}_1(\lambda )R_{21}(\mu ,-\lambda )
{\cal {T}}_2(\mu ) \} .
\nonumber \\
&=&\frac {1}{\tilde {\rho }(-\lambda -\mu )\rho (\lambda -\mu )}
tr_{12}\{ R_{12}(-\lambda ,-\mu )K_1^+(\lambda )
Y_{21}(-\mu -4\eta ,\lambda )K_2^+(\mu )
\} \nonumber \\
&&~~~\{ {\cal {T}}_2(\mu )
R_{12}(\lambda ,-\mu ){\cal {T}}_1(\lambda )R_{21}(-\mu ,-\lambda )
 \} .
\end{eqnarray}
Use again the unitarity relation (17) and 
the cross-unitarity relation (20), and considering the properties of
the function 
\begin{eqnarray}
\rho (\lambda -\mu )&=&\rho (\mu -\lambda ), \nonumber \\
{\tilde {\rho }}(-\lambda -\mu )&=&{\tilde {\rho }}(\lambda +\mu +4\eta ) ,
\end{eqnarray}
we have
\begin{eqnarray}
\cdots &=&\frac {1}{\tilde {\rho }(-\lambda -\mu )}
tr_{12}\{ K_1^+(\lambda )
Y_{21}(-\mu -4\eta ,\lambda )K_2^+(\mu )\} 
\{ {\cal {T}}_2(\mu )
R_{12}(\lambda ,-\mu ){\cal {T}}_1(\lambda )\} .
\nonumber \\
&=&\frac {1}{\tilde {\rho }(-\lambda -\mu )}
tr_{12}\{ Y_{21}^{t_1}(-\mu -4\eta ,\lambda )K_1^+(\lambda )^{t_1}
K_2^+(\mu )\} 
\{ {\cal {T}}_2(\mu ){\cal {T}}_1^{t_1}(\lambda )
R_{12}^{t_1}(\lambda ,-\mu )\} .
\nonumber \\
&=&tr_2K_2^+(\mu ){\cal {T}}_2(\mu )
tr_1K_1^+(\lambda ){\cal {T}}_1(\lambda )
\nonumber \\
&=&t(\mu )t(\lambda )
\end{eqnarray}
Thus we have proved that the transfer matrix $t(\lambda )$
forms a commuting family which gives an infinite set of conserved
quantities.
\begin{eqnarray}
t(\lambda )t(\mu )=t(\mu )t(\lambda )
\end{eqnarray}

One solution of the RE and dual RE correspondes to one transfer matrix.
We have seen that in the proof of the integrability, the RE and dual 
RE are independent of each other. 

By solving directly the RE and the dual RE, we find the following
diagonal solutions to the RE and the dual RE:
\begin{eqnarray}
K(\lambda )=\left( \begin{array}{cc}
1 &0 \\
0& {\frac {\xi (-\lambda )}{\xi (\lambda )}}
{\frac {\sin (\psi +\lambda )}{\sin (\psi -\lambda )}}\end{array} 
\right) ,\\
K^+(\lambda )=\left( \begin{array}{cc}
1 &0 \\
0& {\frac {\xi (\lambda )}{\xi (-\lambda )}}
{\frac {\sin (\psi ^+-\lambda -2\eta )}{\sin (\psi ^++\lambda +2\eta )}}
\end{array} 
\right) ,
\end{eqnarray}
where $\psi $ and $\psi ^+$ are boundary parameters. We noticed that this
diagonal solutions are also the general solutions of the RE (16) and the 
dual RE (21).

Explicitly, the transfer matrix can be written as:
\begin{eqnarray}
t(\lambda )=tr_{\bar {0}}K^+_{\bar {0}}
(\lambda )L_{N{\bar {0}}}(\lambda )
\cdots L_{1{\bar {0}}}(\lambda )K_{\bar {0}}(\lambda )
{\cal {L}}_{1{\bar {0}}}(\lambda )\cdots 
{\cal {L}}_{N{\bar {0}}}(\lambda ).
\end{eqnarray}
Here we denote ${\cal {L}}(\lambda )\equiv L^{-1}(-\lambda )$.
Up to a whole factor, we take the inverse of the $L$ operator as:
\begin{eqnarray}
{\cal {L}}_j(\lambda )=
\left( \begin{array}{cc}
\frac {a''_++b''_+}{2}+\frac {a''_+-b''_+}{2}\sigma _j^z
& c\sigma _j^-\\
c\sigma _j^+ & \frac {a''_-+b''_-}{2}-\frac {a''_--b''_-}{2}\sigma _j^z
\end{array}\right).
\end{eqnarray}
where
\begin{eqnarray}
&&a_+''(\lambda ):b_+''(\lambda ):a_-''(\lambda ):b_-''(\lambda ):c
\nonumber \\
&=&\xi ^{-1}(-\lambda )\sin (\lambda +2\eta )
:\xi (-\lambda )\sin (\lambda )
\nonumber \\
&&: \xi (-\lambda )\sin (\lambda +2\eta ):
\xi ^{-1}(-\lambda )\sin (\lambda ):\sin (2\eta ),
\end{eqnarray}
We define the Hamiltonian for open boundary conditions $H_{OB}$ as follows:
\begin{eqnarray}
H_{OB}&\equiv &-{1\over 2}\sin (2\eta )
\frac {d}{d\lambda }\ln t(\lambda )|_{\lambda =0}
\nonumber \\
&=&-\sum _{j=1}^{N-1}H_{j,j+1}-{1\over 2}\sin (2\eta )K_1'(0)
-\frac {tr_{\bar {0}}K^+_{\bar {0}}(0)L'_{N{\bar {0}}}(0)P_{N{\bar {0}}}}
{trK^+(0)}.
\end{eqnarray}
Here 
\begin{eqnarray}
H_{j,j+1}=L'_{j,j+1}(0)P_{j,j+1}=P_{j,j+1}{\cal {L}}'(0).
\end{eqnarray}
And the relation
\begin{eqnarray}
tr_{\bar {0}}K^+_{\bar {0}}(0)L'_{N{\bar {0}}}(0)P_{N{\bar {0}}}
=tr_{\bar {0}}K^+_{\bar {0}}(0)P_{N{\bar {0}}}(0){\cal {L}}'_{N{\bar {0}}}(0)
\end{eqnarray}
has been used.

So, the open Heisenberg $XXZ$ spin chain with an external magnetic field
parallel to $z$ direction is as follows:
\begin{eqnarray}
H_{OB}&=&-\sum _{j=1}^{N-1}\left( J(\sigma _j^+\sigma _{j+1}^-
+\sigma _j^-\sigma _{j+1}^+)-{V\over 4}
\sigma _j^z\sigma _{j+1}^z\right)
+{{W+V}\over 2}\sum _{j=1}^{N-1}\sigma _j^z
\nonumber \\
&&+{1\over 2}\sin (2\eta )\frac {\cos \psi }{\sin \psi }\sigma _1^z
+{1\over 2}\left( V+W-\sin (2\eta )\frac {\cos \psi ^+}
{\sin \psi ^+}\right) \sigma _N^z
+cost..
\end{eqnarray}

\section{Algebraic Bethe ansatz method}
In this section, we will use the algebraic Bethe ansatz method to 
obtain the eigenvalues and eigenvectors of the transfer matrix
with open boundary conditions. As mentioned above, the 
double-row-monodromy matrix is defined as 
${\cal {T}}(\lambda ) =T(\lambda )$
$\times K(\lambda )T^{-1}(-\lambda )$, and
we denote it as follows:
\begin{eqnarray}
{\cal {T}}(\lambda )&=&\left( \begin{array}{cc}
{\cal {A}}(\lambda )&{\cal {B}}(\lambda )\\ 
{\cal {C}}(\lambda ) &{\cal {D}}(\lambda )\end{array}\right)
=T(\lambda )K(\lambda )T^{-1}(-\lambda )
\nonumber \\
&=&\left( \begin{array}{cc}
A(\lambda )&B(\lambda )\\ 
C(\lambda ) &D(\lambda )\end{array}\right)
\left( \begin{array}{cc}
1 &0 \\
0& {\frac {\xi (-\lambda )}{\xi (\lambda )}}
{\frac {\sin (\psi +\lambda )}{\sin (\psi -\lambda )}}\end{array} 
\right) 
\left( \begin{array}{cc}
{\bar {A}}(-\lambda )&{\bar {B}}(-\lambda )\\ 
{\bar {C}}(-\lambda ) &{\bar {D}}(-\lambda )\end{array}\right)
\end{eqnarray}

We know that one can find easily the pseudovacuum state $|0>$, and
acting the elements of the monodromy matrix $T(\lambda )$ and
$T^{-1}(-\lambda )$ on this pseudovacuum state, we have
\begin{eqnarray}
&&C(\lambda )|0>=0, ~~~\bar {C}(-\lambda )|0>=0, 
\nonumber \\
&&B(\lambda )|0>\not= 0, ~~~\bar {B}(-\lambda )|0>\not= 0,
\nonumber \\
&&A(\lambda )|0>=\alpha (\lambda )|0>, ~~~\bar {A}(-\lambda )
=\bar {\alpha }(-\lambda )|0>,
\nonumber \\
&&D(\lambda )|0>=\delta (\lambda )|0>, ~~~\bar {D}(-\lambda )
=\bar {\delta }(-\lambda )|0>, 
\end{eqnarray}
where
\begin{eqnarray}
\alpha (\lambda )&=&\left( \xi (\lambda )\sin (\lambda +2\eta )\right) ^N,
\nonumber \\
\delta (\lambda )&=&\left( \xi (\lambda )\sin (\lambda )\right) ^N,
\nonumber \\
\bar {\alpha }(-\lambda )&=&
\left( \xi ^{-1}(-\lambda )\sin (\lambda +2\eta )\right) ^N,
\nonumber \\
\bar {\delta }(-\lambda )
&=&\left( \xi ^{-1}(-\lambda )\sin (\lambda )\right) ^N.
\end{eqnarray}
From the Yang-Baxter relation (13), one can obtain
\begin{eqnarray}
T_2^{-1}(\mu )R_{12}(\lambda ,\mu)T_1(\lambda )
=T_1(\lambda )R_{12}(\lambda ,\mu)T_2^{-1}(\mu ).
\end{eqnarray}
Let $\mu =-\lambda $, use the result $C(\lambda )|0>=0$, we have
\begin{eqnarray}
C(\lambda )\bar {B}(-\lambda )|0>&=&
\frac {\xi (-\lambda )}{\xi (\lambda )}
\frac {\sin (2\eta )}{\sin (2\lambda +2\eta )}
\bar {A}(-\lambda )A(\lambda )|0>
\nonumber \\
&&-\frac {\xi (-\lambda )}{\xi (\lambda )}
\frac {\sin (2\eta )}{\sin (2\lambda +2\eta )}
D(\lambda )\bar {D}(-\lambda )|0>.
\end{eqnarray}
For convenience, we introduce here a transformation 
\begin{eqnarray}
{\cal {D}}(\lambda )=\tilde {\cal {D}}(\lambda )
+\frac {\sin (2\eta )}{\sin (2\lambda +2\eta )}
\frac {\xi (-\lambda )}{\xi (\lambda )}  {\cal {A}}(\lambda ).
\end{eqnarray}
Then for double-row-monodromy matrix ${\cal {T}}$, acting its elements
on the pseudovacuum state, and considering the above transformation,
we have
\begin{eqnarray}
&&{\cal {C}}(\lambda )|0>=0, ~~~{\cal {B}}(\lambda )|0>\not =0,
\nonumber \\
&&{\cal {A}}(\lambda )|0>=\alpha (\lambda )\bar {\alpha }(-\lambda )|0>,  
\nonumber \\
&&\tilde {\cal {D}}(\lambda )|0>=
\frac {\xi (-\lambda )}{\xi (\lambda )}
\frac {\sin (2\lambda )\sin (\psi +\lambda +2\eta )}
{\sin (2\lambda +2\eta )\sin (\psi -\lambda )}
\delta (\lambda )\bar {\delta }(-\lambda )|0>.
\end{eqnarray}

With the help of the transformation (43), the transfer matrix for
open boundary conditions can be written as:
\begin{eqnarray}
t(\lambda )&=&{\cal {A}}(\lambda )
+\frac {\xi (\lambda )}{\xi (-\lambda )}
\frac {\sin (\psi ^+-\lambda -2\eta )}
{\sin (\psi ^++\lambda +2\eta )}{\cal {D}}(\lambda )
\nonumber \\
&=&
\frac {\sin (2\lambda +4\eta )\sin (\psi ^++\lambda )}
{\sin (2\lambda +2\eta )\sin (\psi ^++\lambda +2\eta )}{\cal {A}}(\lambda )
+\frac {\xi (\lambda )}{\xi (-\lambda )}
\frac {\sin (\psi ^+-\lambda -2\eta )}
{\sin (\psi ^++\lambda +2\eta )}\tilde {\cal {D}}(\lambda ).
\nonumber \\
\end{eqnarray}

We know that the double-row-monodromy matrix ${\cal {T}}$ satisfy
the RE, from the relation (23),
we have the commutation relations:
\begin{eqnarray}
{\cal {A}}(\lambda ){\cal {B}}(\mu )
&=&\frac {\sin (\lambda -\mu -2\eta )\sin (\lambda +\mu )}
{\sin (\lambda -\mu )\sin (\lambda +\mu +2\eta )}
\frac {\xi ^2(-\lambda )}{\xi ^2(\lambda )}
{\cal {B}}(\mu ){\cal {A}}(\lambda )\nonumber \\
&&+\frac {\sin (2\eta )\sin (2\mu )}{\sin (\lambda -\mu )
\sin (2\mu +2\eta )}
\frac {\xi (-\lambda )\xi (-\mu )}{\xi ^2(\mu )}
{\cal {B}}(\lambda ){\cal {A}}(\mu )\nonumber \\
&&-\frac {\sin (2\eta )}{\sin (\lambda +\mu +2\eta )}
\frac {\xi (-\lambda )}{\xi (\mu )}
{\cal {B}}(\lambda )\tilde {\cal {D}}(\mu )
\end{eqnarray}
\begin{eqnarray}
\tilde {\cal {D}}(\lambda ){\cal {B}}(\mu )&=&
\frac {\sin (\lambda -\mu +2\eta )\sin (\lambda +\mu +4\eta )}
{\sin (\lambda -\mu )\sin (\lambda +\mu +2\eta )}
\frac {\xi ^2(-\lambda )}{\xi ^2(\lambda )}
{\cal {B}}(\mu )\tilde {\cal {D}}(\lambda )\nonumber \\
&&-\frac {\sin (2\eta )\sin (2\lambda +4\eta )}
{\sin (\lambda -\mu )\sin (2\lambda +2\eta )}
\frac {\xi ^2(-\lambda )}{\xi (\lambda )\xi (\mu )}
{\cal {B}}(\lambda )\tilde {\cal {D}}(\mu )\nonumber \\
&&+\frac {\sin (2\eta )\sin (2\mu )\sin (2\lambda +4\eta )}
{\sin (2\lambda +2\eta )\sin (2\mu +2\eta )\sin (\lambda +\mu +2\eta )}
\frac {\xi ^2(-\lambda )\xi (-\mu )}{\xi (\lambda )\xi ^2(\mu )}
{\cal {B}}(\lambda ){\cal {A}}(\mu ),
\nonumber \\
{\cal {B}}(\lambda ){\cal {B}}(\mu )&=&
\frac {\xi ^2(\mu )\xi ^2(-\lambda )}
{\xi ^2(\lambda )\xi ^2(-\mu )}{\cal {B}}(\mu )
{\cal {B}}(\lambda ).
\end{eqnarray}
 
Using the standard algebraic Bethe ansatz method, we assume the
eigenvectors take the form
\begin{eqnarray}
|\mu _1, \cdots ,\mu _n>=
{\cal {B}}(\mu _1)\cdots {\cal {B}}(\mu _n)F(\mu _1,\cdots ,\mu _n)|0>,
\end{eqnarray}
where $F$ is a non-vanishing function. Acting the transfer matrix
on this eigenvector, we obtain the eigenvalues of the transfer matrix
as:
\begin{eqnarray}
&&\Lambda (\lambda ,\mu _1, \cdots ,\mu _n)
\nonumber \\
&=&
\frac {\sin (2\lambda +4\eta )\sin (\psi ^++\lambda )}
{\sin (2\lambda +2\eta )\sin (\psi ^++\lambda +2\eta )}
\nonumber \\
&&\times \alpha (\lambda )\bar {\alpha }(-\lambda )
\prod _{i=1}^n
\left\{ \frac {\sin (\lambda -\mu _i-2\eta )\sin (\lambda +\mu _i)}
{\sin (\lambda -\mu _i)\sin (\lambda +\mu _i+2\eta )}
\frac {\xi ^2(-\lambda )}{\xi ^2(\lambda )}
\right\}
\nonumber \\
&+&
\frac {\sin (\psi ^+-\lambda -2\eta )}
{\sin (\psi ^++\lambda +2\eta )}
\frac {\sin (2\lambda )\sin (\psi +\lambda +2\eta )}
{\sin (2\lambda +2\eta )\sin (\psi -\lambda )}
\delta (\lambda )\bar {\delta }(-\lambda )
\nonumber \\
&&\times \prod _{i=1}^n
\left\{ \frac {\sin (\lambda -\mu _i+2\eta )\sin (\lambda +\mu _i+4\eta )}
{\sin (\lambda -\mu _i)\sin (\lambda +\mu _i+2\eta )}
\frac {\xi ^2(-\lambda )}{\xi ^2(\lambda )}
\right\} ,
\end{eqnarray}
the parameters
$\mu _i, ~~i=1, \cdots ,n$ should satisfy the following Bethe ansatz
equations:
\begin{eqnarray}
\frac {\sin (\mu _j+2\eta )^{2N}}
{\sin (\mu _j)^{2N}}
&=&
\frac {\sin (\psi ^+-\mu _j-2\eta )}
{\sin (\psi ^++\mu _j)}
\frac {\sin (\psi +\mu _j+2\eta )}
{\sin (\psi -\mu _j )}
\nonumber \\
&&\prod _{i=1, i\not= j}^n
\left\{ \frac {\sin (\mu _j-\mu _i+2\eta )\sin (\mu _j+\mu _i+4\eta )}
{\sin (\mu _j-\mu _i-2\eta )\sin (\mu _j+\mu _i)}
\right\},
\nonumber \\
&&j=1, 2, \cdots, n.
\end{eqnarray}
The Bethe ansatz equations ensure that the unwanted terms vanish. On
the other hand, the Bethe ansatz equations also ensure that the 
eigenvalues of the transfer matrix are entire functions.

From the definition of the Hamiltonian (34), we thus can obtain the
eigenvalues $E$ of the Hamiltonian
\begin{eqnarray}
E&=&-N\cos (2\eta )-\sum _{i=1}^n\frac {\sin ^2(2\eta )}
{\sin (\mu _j)\sin (\mu _j+2\eta )}
+(2n-N)\sin (2\eta )\tan (\alpha )
\nonumber \\
&&-\frac {\sin ^2(2\eta )}{2\sin (\psi ^+)\sin (\psi ^++2\eta )}
+cost.
\end{eqnarray}

\section{Integrable model for fermion case}
In this section, we will study the corresponding fermion model.
The Lax representation for the Hamiltonian (1) is found to be [13]
\begin{eqnarray}
L_j^F(\lambda )=\left(
\begin{array}{cc}
b'_+-(b'_+-ia'_+)n_j &ca_j\\
-i ca_j^{\dagger } &a'_--(a'_-+ib'_-)n_j
\end{array}\right) .
\end{eqnarray}
Here $a'_{\pm }, b'_{\pm }$ and $c$ have already be defined in the
above sections.
As for the spin model, we first write out the 
graded Yang-Baxter relation,
\begin{equation}
R_{12}^F(\lambda ,\mu)L_1^F(\lambda )
L_2^F(\mu )=L_2^F(\mu )L_1^F(\lambda )R_{12}^F(\lambda ,\mu),
\end{equation}
we should notice that the meaning of some notations 
here is different from that of
spin models presented above.
Here 
\begin{eqnarray}
L_1^F(\lambda )=L^F(\lambda )\otimes _s1,
\nonumber \\
L_2^F(\lambda )=1\otimes _sL^F(\lambda ).
\end{eqnarray}
$\otimes _s$ denotes the super tensor product
\begin{eqnarray}
[A\otimes _sB]_{ij,kl}=(-1)^{p(j)[p(i)+p(k)]}A_{ik}B_{jl}
\end{eqnarray}
with parity $p(1)=0,~ p(2)=1$. 
The R-matrix for the fermion model is defined as:
\begin{eqnarray}
R_{12}^F(\lambda ,\mu )
\equiv \left(
\begin{array}{cccc}
a_+(\lambda ,\mu ) & 0& 0& 0\\
0 &-ib_-(\lambda ,\mu )& c(\lambda ,\mu )& 0\\
0 &c(\lambda ,\mu )& ib_+(\lambda ,\mu )& 0\\
0& 0& 0& -a_-(\lambda ,\mu )
\end{array} \right) .
\end{eqnarray}
We can prove that this R-matrix satisfy the following
unitarity and cross-unitarity relations
\begin{eqnarray}
&&R_{12}^F(\lambda ,\mu )R_{21}^F(\mu ,\lambda )
=\sin (2\eta +\lambda -\mu )\sin (2\eta -\lambda +\mu )
=\rho (\lambda -\mu ),
\\
&&{X_{12}^F}^{st_1}(\lambda ,\mu +4\eta -\pi )
{R_{21}^F}^{st_1}(\mu ,\lambda )
=\sin (\lambda -\mu )\sin (\lambda -\mu -4\eta )
=-\tilde {\rho }(\lambda -\mu ),
\nonumber \\
&&{R_{12}^F}^{st_1}(\lambda ,\mu )
{Y_{21}^F}^{st_1}(\mu -4\eta +\pi ,\lambda )
=-\tilde {\rho }(\lambda -\mu +4\eta ),
\end{eqnarray}
where we have used the notations:
\begin{eqnarray}
&&X_{12}^F(\lambda ,\mu )\equiv
\frac {\xi (\mu )}{\xi (\mu -4\eta +\pi )}M_1^F(\mu )
R_{12}^F(\lambda ,\mu )M_1^F(\mu ),
\nonumber \\
&&Y_{12}^F(\lambda ,\mu )=\frac {\xi (\lambda +4\eta -\pi )}
{\xi (\lambda )}
M_2^F(\lambda )R_{12}^F(\lambda , \mu )M_2^F(\lambda ).
\nonumber \\
&&M^F(\lambda )=\left( \begin{array}{cc}
1 & 0\\ 
0 &\xi (\lambda -4\eta +\pi )\xi ^{-1}(\lambda )\end{array}\right) .
\end{eqnarray}
Here, 
\begin{eqnarray}
R_{21}^F=P^FR_{12}^FP^F
\end{eqnarray}
$P^F$ is super permutation operator which is defined as:
\begin{eqnarray}
P^F_{ij,kl}=(-1)^{p(i)p(j)}\delta _{il}\delta _{jk} 
\end{eqnarray}
$st$ means super transposition defined as
\begin{eqnarray}
\left( \begin{array}{cc}
A&B\\ C&D\end{array}
\right) ^{st}=
\left( \begin{array}{cc}
A&-C\\ B&D\end{array}\right) ^{st}.
\end{eqnarray}
We can also write the RE and dual RE for fermion models, here 
we take the solutions of RE and dual RE $K^F$ and ${K^+}^F$ to 
be diagonal. It is trick to deal with non-diagonal K-matrix, because
there are grassmann odd numbers in the non-digonal positions.
One can write the graded RE and dual graded RE for fermion model as:
\begin{eqnarray}
R_{12}^F(\lambda ,\mu )K_1^F(\lambda )R_{21}^F(\mu ,-\lambda )K_2^F(\mu ) 
=K_2(\mu )^FR_{12}^F(\lambda ,-\mu )K_1^F(\lambda )R_{21}^F(-\mu ,-\lambda ).
\end{eqnarray}
\begin{eqnarray}
R_{12}^F(-\lambda ,-\mu ){K^+_1}^F(\lambda )
Y_{21}^F(-\mu -4\eta+\pi  ,\lambda ){K_2^+}^F(\mu )
\nonumber 
\\
={K_2^+}^F(\mu )X_{12}^F(-\lambda ,\mu +4\eta -\pi )K_1^F(\lambda )
R_{21}^F(\mu ,\lambda ).
\end{eqnarray}
By solving the graded RE and the dual graded RE, 
we find respectively the solutions of the graded RE and dual graded RE as: 
\begin{eqnarray}
K^F(\lambda )=\left( \begin{array}{cc}
{\frac {\xi (\lambda )}{\xi (-\lambda )}}
{\frac {\sin (\psi -\lambda )}{\sin (\psi +\lambda )}}
&0 \\
0&1 
\end{array} 
\right) ,\\
{K^+}^F(\lambda )=\left( \begin{array}{cc}
{\frac {\xi (-\lambda )}{\xi (\lambda )}}
{\frac {\sin (\psi ^++\lambda +2\eta )}{\sin (\psi ^+-\lambda -2\eta )}}
&0 \\
0&-1 
\end{array} 
\right) .
\end{eqnarray}
One can find $K^F$ is similar as the spin model, but there appear 
a $-$ in ${K^+}^F$ compared with spin model. 
The transfer matrix for fermion models is defined as:
\begin{eqnarray}
t^F(\lambda )=str{K^+}^F(\lambda ){\cal {T}}^F(\lambda )
\end{eqnarray}
Here, $str$ means super trace defined as $str A=(-1)^{p(i)}A_{ii}$.
And
\begin{eqnarray}
{\cal {T}}^F(\lambda )
=L_N^F(\lambda )\cdots L_1^F(\lambda )K^F(\lambda )
{\cal {L}}^F_1(\lambda )\cdots {{\cal {L}}_N^F}(\lambda )
\end{eqnarray}
satisfy the graded RE. We have used the notation
${\cal {L}}^F(\lambda )\equiv {L^F}^{-1}(-\lambda )$, which is defined as:
\begin{eqnarray}
{\cal {L}}_j^F(\lambda )=\left(
\begin{array}{cc}
-b''_++(b''_++ia''_+)n_j &-ica_j\\
-ca_j^{\dagger } &-a''_-+(a''_--ib''_-)n_j
\end{array}\right) .
\end{eqnarray}
The definition for  
$a''_{\pm }, b''_{\pm }$ and $c$ can be found in the above sections.
The Hamiltonian ${\cal {H}}_{OB}$
for small-polaron model with open boundary conditions
is as follows:
\begin{eqnarray}
{\cal {H}}_{OB}&=&W\sum _{j=1}^{N-1}n_j
-J\sum _{j=1}^{N-1}(a_j^{\dagger }a_{j+1}+a_{j+1}^{\dagger }a_j)
+V\sum _{j=1}^{N-1}n_jn_{j+1},
\nonumber \\
&&+\left( {V\over 2}+\sin (2\eta )\frac {\cos \psi }{\sin \psi }\right) n_1
+\left( {V\over 2}+W-\sin (2\eta )\frac {\cos \psi ^+}
{\sin \psi ^+}\right) n_N
+cost.
\nonumber \\
\end{eqnarray}

The commuting of the transfer matrix for fermion model 
with open boundary conditions
can be proved similarly as that for 
the spin model, here we should notice that 
in the proof of the integrability, besides the super transposition
$st$,
we also need a inverse of the super transposition $\bar {st}$
with ${A^{st}}^{\bar {st}}=A$.

We denote 
\begin{eqnarray}
{\cal {T}}^F(\lambda )&=&\left( \begin{array}{cc}
{\cal {A}}^F(\lambda )&{\cal {B}}^F(\lambda )\\ 
{\cal {C}}^F(\lambda ) &{\cal {D}}^F(\lambda )\end{array}\right) ,
\end{eqnarray}
thus we can write the transfer matrix explicitly as 
\begin{eqnarray}
t^F(\lambda )&=&
{\frac {\xi (-\lambda )}{\xi (\lambda )}}
{\frac {\sin (\psi ^++\lambda +2\eta )}{\sin (\psi ^+-\lambda -2\eta )}}
{\cal {A}}^F(\lambda )
+{\cal {D}}^F(\lambda )
\nonumber \\
&=&{\frac {\xi (-\lambda )}{\xi (\lambda )}}
{\frac {\sin (\psi ^++\lambda +2\eta )}{\sin (\psi ^+-\lambda -2\eta )}}
\tilde {\cal {A}}^F(\lambda )
\nonumber \\
&&+\frac {\sin (2\lambda +4\eta )\sin (\psi ^+-\lambda )}
{\sin (2\lambda +2\eta )\sin (\psi ^+-\lambda -2\eta )}
{\cal {D}}^F(\lambda ).
\end{eqnarray}
In above, as for the case of spin model, we have introduced
the following transformation for convenience,
\begin{eqnarray}
{\cal {A}}^F(\lambda )
=\tilde {\cal {A}}^F(\lambda )+
\frac {\xi (\lambda )}{\xi (-\lambda )}
\frac {\sin (2\eta )}{\sin (2\lambda +2\eta )}
{\cal {D}}^F(\lambda ).
\end{eqnarray}

Following the standard algebraic Bethe ansatz method, we define the
pseudovacuum state for fermion model as $|0>^F$ with $a_j|0>^F=0$.
One can find
\begin{eqnarray}
&&\tilde {\cal {A}}^F(\lambda )|0>^F=
\frac {\xi (\lambda )}{\xi (-\lambda )}
\frac {\sin (2\lambda )\sin (\psi -\lambda -2\eta )}
{\sin (2\lambda +2\eta )\sin (\psi +\lambda )}\alpha ^F(\lambda )|0>^F,
\nonumber \\
&&{\cal {D}}^F(\lambda )|0>^F=\delta ^F(\lambda )|0>^F,
\nonumber \\
&&{\cal {B}}^F(\lambda )|0>^F=0,~~~
{\cal {C}}^F(\lambda )|0>^F\not =0.
\end{eqnarray}
Where we have
\begin{eqnarray}
\alpha ^F(\lambda )=
\left( -\frac {\xi (-\lambda )}{\xi (\lambda )}
\right) ^N\sin ^{2N}(\lambda ),
\delta ^F(\lambda )=\left( -\frac {\xi (-\lambda )}{\xi (\lambda )}
\right) ^N\sin ^{2N}(\lambda +2\eta ).
\end{eqnarray}
Because the double-row-monodromy matrix for fermion model satisfy
the graded RE, we can obtain the commutation relations which are
necessary for the algebraic Bethe ansatz method.
\begin{eqnarray}
{\cal {C}}^F(\lambda ){\cal {C}}^F(\mu )&=&
\frac {\xi ^2(\lambda )\xi ^2(-\mu )}{\xi ^2(\mu )\xi ^2(-\lambda )}
{\cal {C}}^F(\mu )
{\cal {C}}^F(\lambda ),
\nonumber \\
\tilde {\cal {A}}^F(\lambda ){\cal {C}}^F(\mu )&=&
\frac {\sin (\lambda -\mu +2\eta )\sin (\lambda +\mu +4\eta )}
{\sin (\lambda -\mu )\sin (\lambda +\mu +2\eta )}
\frac {\xi ^2(\lambda )}{\xi ^2(-\lambda )}
{\cal {C}}^F(\mu )\tilde {\cal {A}}^F(\lambda )\nonumber \\
&&-\frac {\sin (2\eta )\sin (2\lambda +4\eta )}
{\sin (\lambda -\mu )\sin (2\lambda +2\eta )}
\frac {\xi (\lambda )\xi (\mu )}{\xi ^2(-\lambda )}
{\cal {C}}^F(\lambda )\tilde {\cal {A}}^F(\mu )\nonumber \\
&&+\frac {\sin (2\eta )\sin (2\mu )\sin (2\lambda +4\eta )}
{\sin (2\lambda +2\eta )\sin (2\mu +2\eta )\sin (\lambda +\mu +2\eta )}
\nonumber \\
&&\times \frac {\xi (\lambda )\xi ^2(\mu )}{\xi ^2(-\lambda )\xi (-\mu )}
{\cal {C}}^F(\lambda ){\cal {D}}^F(\mu ),
\nonumber \\
{\cal {D}}^F(\lambda ){\cal {C}}^F(\mu )
&=&\frac {\sin (\lambda -\mu -2\eta )\sin (\lambda +\mu )}
{\sin (\lambda -\mu )\sin (\lambda +\mu +2\eta )}
\frac {\xi ^2(\lambda )}{\xi ^2(-\lambda )}
{\cal {C}}^F(\mu ){\cal {D}}^F(\lambda )\nonumber \\
&&+\frac {\sin (2\eta )\sin (2\mu )}{\sin (\lambda -\mu )
\sin (2\mu +2\eta )}
\frac {\xi ^2(\mu )}{\xi (-\lambda )\xi (-\mu )}
{\cal {C}}^F(\lambda ){\cal {D}}^F(\mu )\nonumber \\
&&-\frac {\sin (2\eta )}{\sin (\lambda +\mu +2\eta )}
\frac {\xi (\mu )}{\xi (-\lambda )}
{\cal {C}}^F(\lambda )\tilde {\cal {A}}^F(\mu ).
\end{eqnarray}
Those commutation relations are only slightly different from the
relations for spin case. Assume the eigenvectors take the form
${\cal {C}}^F(\mu _1)\cdots {\cal {C}}^F(\mu _n)
F(\mu _1,\cdots ,\mu _n)|0>^F$, where $F(\mu _1, \cdots \mu _n)$
are non-vanishing functions. We can find
the eigenvalues for the fermion transfer matrix with open boundary
conditions are:
\begin{eqnarray}
&&\Lambda ^F(\lambda ,\mu _1, \cdots ,\mu _n)
\nonumber \\
&=&{\frac {\sin (\psi ^++\lambda +2\eta )}{\sin (\psi ^+-\lambda -2\eta )}}
\frac {\sin (2\lambda )\sin (\psi -\lambda -2\eta )}
{\sin (2\lambda +2\eta )\sin (\psi +\lambda )}\alpha ^F(\lambda )
\nonumber \\
&&\times \prod _{i=1}^n
\left\{ \frac {\sin (\lambda -\mu _i+2\eta )\sin (\lambda +\mu _i+4\eta )}
{\sin (\lambda -\mu _i)\sin (\lambda +\mu _i+2\eta )}
\frac {\xi ^2(\lambda )}{\xi ^2(-\lambda )}
\right\} 
\nonumber \\
&&+\frac {\sin (2\lambda +4\eta )\sin (\psi ^+-\lambda )}
{\sin (2\lambda +2\eta )\sin (\psi ^+-\lambda -2\eta )}
\delta ^F(\lambda )
\nonumber \\
&&\times
\prod _{i=1}^n
\left\{ \frac {\sin (\lambda -\mu _i-2\eta )\sin (\lambda +\mu _i)}
{\sin (\lambda -\mu _i)\sin (\lambda +\mu _i+2\eta )}
\frac {\xi ^2(\lambda )}{\xi ^2(-\lambda )}
\right\}.
\end{eqnarray}
The parameters $\mu _1,\cdots ,\mu _n$
are restricted by the Bethe ansatz equations:
\begin{eqnarray}
\frac {\sin ^{2N}(\mu _j+2\eta )}
{\sin ^{2N}(\mu _j)}
&=&
\frac {\sin (\psi ^++\mu _j+2\eta )}
{\sin (\psi ^+-\mu _j)}
\frac {\sin (\psi -\mu _j-2\eta )}
{\sin (\psi +\mu _j )}
\nonumber \\
&&\prod _{i=1, i\not= j}^n
\left\{ \frac {\sin (\mu _j-\mu _i+2\eta )\sin (\mu _j+\mu _i+4\eta )}
{\sin (\mu _j-\mu _i-2\eta )\sin (\mu _j+\mu _i)}
\right\},
\nonumber \\
&&j=1, 2, \cdots, n.
\end{eqnarray}
The eigenvalue of the Hamiltonian for fermion model is:
\begin{eqnarray}
E^F&=&
-N\cos (2\eta )-\sum _{i=1}^n\frac {\sin ^2(2\eta )}
{\sin (\mu _j)\sin (\mu _j+2\eta )}
+(2n-N)\sin (2\eta )\tan (\alpha )
\nonumber \\
&&-\frac {\sin ^2(2\eta )}{2\sin (\psi ^+)\sin (\psi ^+-2\eta )}
+cost.
\end{eqnarray}

\section{Summary and discussions}
We formulated in this paper the RE and the dual RE for small-polaron model
whose spin chain equivalent is the Heisenberg $XXZ$ chain with an external
magnetic
field parallel to $z$ direction. We found solutions to the RE and dual
RE for both spin and fermion models.
Using the algebraic Bethe ansatz method, we find the eigenvalues of the
transfer matrix with open boundary conditions.

It is interesting to use some results of this paper to calculate some
physical quantities, such as the surface
free ennergy and finite-size corrections,
by using the thermodynamic Bethe ansatz method.
The Lax pair method for open boundary conditions for this model is also
an interesting problem.

\vskip 1truecm
{\bf Acknowlegements:} We would like to thank Prof. B.Y.Hou and
Prof. K.J.Shi for useful discussions. This work is supported in part by
the Natural Science Foundation of China.

\end{document}